\documentclass[pra,twocolumn]{revtex4}
\usepackage{amssymb}
\usepackage{amsfonts}
\usepackage{amsmath}
\usepackage{graphicx}

\begin{document}

\title{Superclassical non-Markovian open quantum dynamics}
\author{Adri\'{a}n A. Budini}
\affiliation{Consejo Nacional de Investigaciones Cient\'{\i}ficas y T\'{e}cnicas
(CONICET), Centro At\'{o}mico Bariloche, Avenida E. Bustillo Km 9.5, (8400)
Bariloche, Argentina, and Universidad Tecnol\'{o}gica Nacional (UTN-FRC),
Fanny Newbery 111, (8400) Bariloche, Argentina}
\date{\today }

\begin{abstract}
We characterize a class of superclassical non-Markovian open quantum system
dynamics that are defined by their lack of measurement invasiveness when the
corresponding observable commutates with the pre-measurement state. This
diagonal non-invasiveness guarantees that joint probabilities for
measurement outcomes fulfill classical Kolmogorov consistency conditions.
These features are fulfilled regardless of the previous (measurement) system
history and are valid at arbitrary later times after an arbitrary system
initialization. It is shown that a subclass of depolarizing dynamics, which
are based on a (time-irreversible) non-unitary system-environment coupling,
satisfy the required properties. The relationship with other operational
[Milz \textit{et al.}, Phys. Rev. X \textbf{10}, 041049 (2020)] and
non-operational [Banacki \textit{et al.}, Phys. Rev. A \textbf{107}, 032202
(2023)] notions of classicality in non-Markovian open quantum systems is
studied in detail and exemplified through different examples.
\end{abstract}

\maketitle

\section{Introduction}

The notion of memory, or equivalently non-Markovianity, in open quantum
systems~\cite{breuerbook,vega}\ has been tackled from different
perspectives. Broadly speaking, the contributions of the last decades can be
split in two different lines of research~\cite{notapie}. In \textit{%
non-operational approaches} the presence of memory is determined by taking
into account departures of the system density matrix propagator with respect
to a Lindblad equation. Given that this departure can be defined in quite
different grounds, diverse non-equivalent formalisms have been proposed in
the literature~\cite{BreuerReview,plenioReview}. On the contrary, in \textit{%
operational approaches }to quantum non-Markovianity the system is subjected
to a series of consecutive measurement processes. Consequently, there is a
unique criterion for defining the presence of memory effects in the system
dynamics. In fact, departure of joint outcome probabilities with respect to
a standard Markovian structure determine this feature~\cite{modiNM,CPF}.

Even when restricted to open quantum dynamics, the previous complementary
perspectives do not have associated any inherent criterion that allow to
splitting non-Markovian memory effects in quantum or classical. In fact,
different (non-equivalent) definitions of \textit{classicality} has been
recently proposed in the literature~\cite%
{horoNoMark,Markol,garcia,maria,plenioExper,plenioNoMarkov,taranto}.


The notion of environment-to-system backflow of information~\cite%
{BreuerReview} is a cornerstone in the definition of memory effects in
non-operational approaches. In Ref.~\cite{horoNoMark} the authors proposed a
criterion to distinguish classical from quantum information backflows.
Essentially, this non-Markovian effect is classified as classical if the
system propagator can be written as a composition of a (classical)
probability map followed by a system unitary transformation. This
(non-operational) definition of classicality makes sense because in such a
case the role of the environment can be mimicked without involving any
quantum feature.

In general, it is not possible to translate straightforwardly the definition
of (environment-to-system) backflow of information to operational approaches~%
\cite{entropy}. Reasonably, the definition of classicality in measurement
based approaches has been tackled from a different perspective. In Refs.~%
\cite{Markol,garcia,maria,plenioExper,plenioNoMarkov,taranto}, given a fixed
measurement basis, the system is defined as classical (behaves as a
classical one) if Kolmogorov consistency conditions are satisfied, that is,
the joint probability for an arbitrary subset of measurement processes can
be obtained by marginalization of the joint probability of the full
measurement set~\cite{breuerbook}. Equivalently, the dynamics (strictly the
outcome statistics) is classical if (in the chosen basis) a non-selective
measurement~\cite{breuerbook} performed at an arbitrary time is noninvasive,
that is, it does not modify the subsequent measurement statistics.


Classicality in operational approaches has been analyzed in both Markovian
and non-Markovian open quantum dynamics. In the Markovian case~\cite%
{Markol,garcia}, the outcome statistics is classical (Kolmogorov conditions
are fulfilled) when, in the chosen measurement basis, the system does not
develop any quantum coherence. In the non-Markovian regime, the absence of
coherences in the measurement basis does not guaranty classicality. In fact,
in this case the role of the coherences is played instead by the
system-environment ($s$-$e$) quantum discord~\cite{plenioNoMarkov}.
Furthermore, even when the dynamics induces coherent behaviors, or quantum
discord, the measurement statistics can appear as a classical one when the
developed quantum features cannot be detected in the chosen measurement
basis (non-coherence-generating-and detecting maps~\cite{Markol} and
non-discord-generating-and detecting maps~\cite{plenioNoMarkov}
respectively).

In a previous contribution~\cite{DNI}, based on an operational scheme, a
deep relationship between measurement invasiveness (a non-classical feature)
and the presence of memory effects has been established. The operational
scheme consists on performing a set of three consecutive (projective)
measurement processes. The first and last are \textit{arbitrary}, while the
intermediate one is performed in the basis where the pre-measurement system
state is \textit{diagonal}. This last constraint eliminates the possibility
of any measurement invasiveness in the (memoryless) Markovian case.
Therefore, given that classical systems neither quantum systems in absence
of memory effects can violate this diagonal non-invasiveness (DNI),
departures from this condition become an indicator of \textit{non-classical}
(quantum) memory effects. This univocal relationship was established for
open quantum systems whose dynamics is derived from a full unitary
microscopic description or from stochastic Hamiltonian models.

The relation between memory effects and measurement invasiveness (violation
of DNI) is valid in general even when the underlying $s$-$e$ coupling is
non-unitary. 
This modelling applies when the system of interest is a subpart of a larger
one and where the evolution of the total arrangement (between system
measurements) can be approximated by a (bipartite) Lindblad equation~\cite%
{tripartito}. Nevertheless, for these evolutions, even when the previous and
posterior measurement processes are arbitrary, DNI could be fulfilled in
presence of memory effects~\cite{DNI}. The corresponding subclass of system
dynamics that fulfill this property are termed as \textit{superclassical}.
By definition, for these non-Markovian evolutions an intermediate
measurement, performed in the basis where the pre-measurement state is
diagonal, is non-invasive independently of the previous and
\textquotedblleft posterior\textquotedblright\ measurement histories.

The main goal of this contribution is to characterize superclassical
non-Markovian open quantum dynamics. Of particular interest is to compare
this notion of classicality with previous non-operational~\cite{horoNoMark}
and operational approaches~\cite{plenioNoMarkov}. Furthermore, we ask about
which kind of system dynamics could fulfill the corresponding definitions.
The answer to this question involves the family of depolarizing dynamics.
Interestingly, these dynamics play a central role not only in this
quantum-classical border, but also in the modeling of noise sources in
quantum information platforms~\cite%
{HoroEntangled,bose,amosov,cirac,mancini,faniza,learn,chau,self,he} and
related theoretical and experimental problems~\cite%
{sasaki,narang,kim,sudarshan,mexico,cresser,petruccione,banerje}.

The manuscript is organized as follows. In Sec. II we review and compare the
different operational and non-operational notions of classicality, together
with superclassicality. In Sec. III we find non-Markovian continuous-in-time
evolutions that fulfill the different definitions of classicality. In Sec.
IV we study a set of specific representative examples. In Sec. V we provide
the conclusions. Demonstrations and calculus details are provided in the
Appendixes.

\section{Definition of classicality in presence of memory effects}

In this section we review and compare the different notion of classicality
when considering non-Markovian open quantum systems. While each formalism is
self-consistent, similarly to the different definition of non-Markovianity,
they are not equivalent between them.

\subsection{Non-operational approach}

A standard non-operational definition of non-Markovianity is based on the
time-behavior of the distance between two system initial conditions~\cite%
{BreuerReview}. An increasing of this distance is read as an
environment-to-system backflow of information, which indicates the presence
of memory effects. In the approach of Ref.~\cite{horoNoMark}, an information
backflow (non-Markovian memory effect) is considered classical if the system
propagator $\Lambda _{t,0}$ assumes the structure%
\begin{equation}
\rho _{t}=\Lambda _{t,0}[\rho _{0}]=U_{t}\Lambda _{t,0}^{\mathrm{cl}}[\rho
_{0}]U_{t}^{\dag },  \label{JoroClassPropa}
\end{equation}%
where $\rho _{0}$\ is the system initial condition. Furthermore, $U_{t}$ is
an arbitrary (system) unitary transformation while%
\begin{equation}
\Lambda _{t,0}^{\mathrm{cl}}[\rho _{0}]=\sum_{c,c^{\prime }}\Lambda
_{cc^{\prime }}(t)|c\rangle \langle c^{\prime }|\rho _{0}|c^{\prime }\rangle
\langle c|.  \label{ProbaMap}
\end{equation}%
Here, the coefficients $\{\Lambda _{cc^{\prime }}(t)\}$ define a map
(propagator) in the space of classical probability vectors $\{p_{c}\},$ that
is, $p_{c}(t)=\sum_{c^{\prime }}\Lambda _{cc^{\prime }}(t)p_{c^{\prime
}}(0). $ On the other hand, $\{|c\rangle \}$ is a \textquotedblleft
particular basis\textquotedblright\ of the system Hilbert space.
Classicality is also valid when the propagator can be written as a
time-independent convex combination%
\begin{equation}
\Lambda _{t,0}=\sum_{i}p_{i}\Lambda _{t,0}^{(i)}[\rho _{0}],  \label{convex}
\end{equation}%
where each map $\Lambda _{t,0}^{(i)}$ fulfills Eq.~(\ref{JoroClassPropa}).

\subsection{Operational approach}

In operational approaches, the presence of memory effects is sets by
departures of the joint outcome probabilities from a standard Markov
condition~\cite{modiNM,CPF}. Under this notion of non-Markovianity, in the
proposal of Ref.~\cite{plenioNoMarkov} the definition of classicality
involves the bipartite $s$-$e$ propagator, $\mathcal{G}_{t+\tau ,t}.$
Furthermore, it is intrinsically related to a\textit{\ chosen} complete
basis $\{|c\rangle \}$ of system states. The dynamics is classical (in the
chosen basis) if%
\begin{equation}
\bigtriangleup \circ \mathcal{G}_{t+\tau ,t}\circ \bigtriangleup \circ 
\mathcal{G}_{t,0}\circ \bigtriangleup =\bigtriangleup \circ \mathcal{G}%
_{t+\tau ,t}\circ \mathcal{G}_{t,0}\circ \bigtriangleup ,  \label{Fixed}
\end{equation}%
where $\circ $\ denotes composition and $\bigtriangleup \lbrack \bullet
]=\sum_{c}|c\rangle \langle c|\bullet |c\rangle \langle c|$ is a complete
dephasing map. Reading this map as the action of a non-selective
measurement, the equality~(\ref{Fixed}) guaranty the fulfillment of
(classical) Kolmogorov conditions of the corresponding outcome probabilities~%
\cite{plenioNoMarkov}. For simplicity, in the above brief overview we have
taken the initial time $t_{0}=0$ and two posterior arbitrary times $t$ and $%
t+\tau .$

Assuming an initial bipartite state with null discord $\rho
_{0}^{se}=\sum_{c}|c\rangle \langle c|\otimes \sigma _{c},$ where $\{\sigma
_{c}\}$ are environment states, the above definition can be fulfilled when
the bipartite propagator do not create quantum discord in the chosen system
basis. Alternatively, the propagator can create quantum discord but this
discord cannot be detected by means of later measurement processes performed
on the system. These two cases define the
non-discord-generating-and-detecting maps~\cite{plenioNoMarkov}.

\subsection{Superclassicality}

In the operational approach of Ref.~\cite{DNI} non-Markovianity is also
defined from the joint outcome probabilities. Furthermore, it establishes a
depper connection between measurement invasiveness and \textquotedblleft
non-classical\textquotedblright\ memory effects. The main difference with
respect to Ref.~\cite{plenioNoMarkov} is the election of the measurement
processes.

The formalism relies on the DNI property of Markovian dynamics, that is,
these dynamics are unaffected by a projective measurement when the
observable commutates with the pre-measurement state. This property is
independent of the previous system history. Non-Markovian dynamics that
fulfill DNI for arbitrary (previous and \textquotedblleft
posterior\textquotedblright ) measurement histories are termed as \textit{%
superclassical}. In terms of the bipartite propagator $\mathcal{G}_{t+\tau
,t}$ they are defined by the equality%
\begin{equation}
\bigtriangleup _{Z}\circ \mathcal{G}_{t+\tau ,t}\circ \bigtriangleup
_{Y}^{t}\circ \mathcal{G}_{t,0}\circ \bigtriangleup _{X}=\bigtriangleup
_{Z}\circ \mathcal{G}_{t+\tau ,t}\circ \mathcal{G}_{t,0}\circ \bigtriangleup
_{X}.  \label{DeltaSuper}
\end{equation}%
In contrast to Eq.~(\ref{Fixed}), this equality must be valid for arbitrary
dephasing maps $\bigtriangleup _{X}$ and $\bigtriangleup _{Z},$ which
correspond to \textit{arbitrary} measurement processes, $\bigtriangleup
_{M}[\bullet ]=\sum_{m}|m\rangle \langle m|\bullet |m\rangle \langle m|,$
where $m=x,y,z.$ Furthermore, $\bigtriangleup _{Y}^{t}$ is the measurement
dephasing map that leaves invariant the pre-measurement system state $\rho
_{t|X},$ where%
\begin{equation}
\rho _{t|X}\equiv \mathrm{Tr}_{e}(\mathcal{G}_{t,0}[\bigtriangleup _{X}[\rho
_{0}]\otimes \sigma _{0}])=\sum_{c}p_{t}^{c}|c_{t}\rangle \langle c_{t}|.
\label{EigenSystem}
\end{equation}%
The bipartite initial condition is $\rho _{0}^{se}=\rho _{0}\otimes \sigma
_{0}.$ The expression for $\rho _{t|X}$ implies that $\bigtriangleup
_{Y}^{t} $ must to commutates with the eigenbasis $\{|c_{t}\rangle \}.$
Equivalently,%
\begin{equation}
\bigtriangleup _{Y}^{t}[\rho _{t|X}]=\rho _{t|X}.  \label{YtMeaurement}
\end{equation}%
Notice that the basis $\{|c_{t}\rangle \},$ as well as the eigenvalues $%
\{p_{t}^{c}\},$ are in general time-dependent. Furthermore, they depend on
which $X$-measurement process was performed. For notational simplicity this
dependence is omitted.

Due to the arbitrariness of $\bigtriangleup _{X}$ and $\bigtriangleup _{Z},$
it is clear that in general superclassicality is a much stronger condition
than the classicality defined by Eq.~(\ref{Fixed}). Nevertheless, they share
a common feature. In fact, Eq.~(\ref{DeltaSuper}) can be fulfilled with or
without quantum discord generation (see next Section). On other hand, in
specific situations it occurs that the intermediate basis does not depend on
time $\bigtriangleup _{Y}^{t}=\bigtriangleup _{Y},$ and in addition $%
\bigtriangleup _{Y}=\bigtriangleup _{X}=\bigtriangleup _{Z}=\bigtriangleup .$
In such a case, justifying its naming, superclassicality fulfills the
classicality definition~(\ref{Fixed}) for any map $\bigtriangleup .$

\subsection{Quantifying departures from superclassicality and memory effects}

Departure from the equality~(\ref{DeltaSuper}) can be quantified as follows~%
\cite{DNI}. Performing three successive measurement processes, the outcome
join-probability is denoted as $P_{3}(z,y,x),$ where the subindex denote the
number of performed measurement processes. Introducing the marginal
probability%
\begin{equation}
P_{3}(z,x)\equiv \sum_{y}P_{3}(z,y,x),  \label{P3Conjunta}
\end{equation}%
departures from DNI [the equality is not valid in Eq.~(\ref{DeltaSuper})]
can be measured by the (probability) distance~\cite{nielsen}%
\begin{equation}
I(t,\tau )=\sum\nolimits_{zx}|P_{3}(z,x)-P_{2}(z,x)|.  \label{IMeasure}
\end{equation}%
Here, $P_{2}(z,x)$ is the joint probability for the outcomes of the first
and last measurements (performed at times $0$ and $t+\tau )$ when the
intermediate measurement (at time $t)$ is not performed. Superclassicality
correspond to non-Markovian dynamics, $P_{3}(z,y,x)\neq
P_{2}(z|y)P_{2}(y|x)P_{1}(x),$ that in turn fulfills%
\begin{equation}
I(t,\tau )=0\ \ \ \ \Leftrightarrow \ \ \ \ P_{3}(z,x)=P_{2}(z,x).
\label{SuperCondition}
\end{equation}%
Consequently, the Kolmogorov consistence condition $P_{3}(z,x)=P_{2}(z,x)$
is fulfilled in presence of memory effects. We remark that in contrast to
Eq.~(\ref{Fixed}), here this property must be valid for arbitrary $X$- and $%
Z $-measurement processes, while the intermediate $Y$-measurement must leave
invariant the system state.

Using the same scheme based on three measurement processes, memory effects
can be witnessed with a conditional past-future correlation~\cite{CPF},
which reads%
\begin{equation}
C_{pf}(t,\tau )=\sum\nolimits_{zx}zx[P_{3}(z,x|y)-P_{3}(z|y)P_{2}(x|y)],
\label{CPFDefinition}
\end{equation}%
where $P_{3}(z,x|y)=P_{3}(z,y,x)/\sum_{x}P_{2}(y,x).$ Markovianity $%
[P_{3}(z,y,x)=P_{2}(z|y)P_{2}(y|x)P_{1}(x)]$ translates to the condition $%
C_{pf}(t,\tau )=0,$ which must be valid for arbitrary $X$-, $Y$-, and $Z$%
-measurement processes. Complementarily, if $C_{pf}(t,\tau )\neq 0$ for some
measurement set, the dynamics is non-Markovian.

Both $I(t,\tau )$ and $C_{pf}(t,\tau )$ provide simple measures of
measurement invasiveness and memory effects respectively. They are
explicitly calculated in the examples of Sec.~\ref{SecExamples}.

\section{Continuous-time dynamics that fulfill classicality}

Given the previous approaches to classicality in presence of memory effects,
here we search which kind of continuous-in-time system dynamics could
fulfill the corresponding definitions.

\subsection{Non-operational approach}

Fixing the time $t,$ a vast class of maps $\rho _{0}\rightarrow \rho _{t}$
can fulfill the classicality conditions Eqs.~(\ref{JoroClassPropa}) and~(\ref%
{ProbaMap}). Nevertheless, here we search dynamics that fulfill classicality
at any time. On the other hand, given a time-evolution, classicality could
be fulfilled in a given basis $\{|c\rangle \}$ or by restricting the initial
condition $\rho _{0}.$ In the present study we disregard these kinds of
\textquotedblleft accidental classicality.\textquotedblright\ Hence, we
demand the validity of Eqs.~(\ref{JoroClassPropa}) and~(\ref{ProbaMap}) at
any time and for arbitrary basis and input states. The motivation for these
requirements is to find dynamical conditions under which the different
notion of classicality could coincide.

Under the above conditions, it is simple to realize that the map $\Lambda
_{t,0}^{\mathrm{cl}}$ only change the eigenvalues of the input state while
maintains the basis where it is a diagonal operator. As demonstrated in
Appendix~\ref{UniVsDe}, this property is only fulfilled by \textit{%
depolarizing maps}, whose structure is%
\begin{equation}
\mathcal{D}_{\lambda }[\rho ]\equiv \lambda \rho +(1-\lambda )\frac{\mathrm{I%
}_{s}}{d}.  \label{DepolMap}
\end{equation}%
Here, $\mathrm{I}_{s}$ is the identity matrix while $d$ is the dimension of
the system Hilbert space. Furthermore, $\lambda $ is an arbitrary weight, $%
0\leq \lambda \leq 1$ \cite{nielsen}. Thus, we write Eq.~(\ref%
{JoroClassPropa}) as%
\begin{equation}
\rho _{t}=U_{t}\Big{[}\lambda _{t}\rho _{0}+(1-\lambda _{t})\frac{\mathrm{I}%
_{s}}{d}\Big{]}U_{t}^{\dagger },  \label{RhoSistema}
\end{equation}%
where $\lambda _{t}$ is a time-dependent weight with $\lambda _{0}=1.$
Consequently, the coefficients of the classical map $\Lambda _{t,0}^{\mathrm{%
cl}}$ [Eq.~(\ref{ProbaMap})] read $\Lambda _{cc^{\prime }}(t)=\delta
_{cc^{\prime }}\lambda _{t}+(1-\lambda _{t})/d,$ where $\delta _{cc^{\prime
}}$ is the Kronecker-delta function.

From Eq.~(\ref{RhoSistema}), the evolution of the system density matrix can
be written as%
\begin{equation}
\frac{d\rho _{t}}{dt}=-i[H_{s},\rho _{t}]+\gamma _{t}(\mathcal{D}_{\mathrm{I}%
}[\rho _{t}]-\rho _{t}).  \label{RhoSEvolution}
\end{equation}%
The superoperator action is $\mathcal{D}_{\mathrm{I}}[\rho ]\equiv \mathrm{I}%
_{s}/d.$ It can always be written as%
\begin{equation}
\mathcal{D}_{\mathrm{I}}[\rho ]=\frac{\mathrm{I}_{s}}{d}=\sum_{k}W_{k}\rho
W_{k}^{\dagger },  \label{HWeyl}
\end{equation}%
where the set of Heisenberg-Weyl operators $\{W_{k}\}$~\cite{wilde} define
its Kraus representation. Furthermore, for simplicity, we assumed a
time-independent system Hamiltonian, $U_{t}=\exp [-itH_{s}].$ The
time-dependent rate $\gamma _{t}$ is%
\begin{equation}
\gamma _{t}=-\frac{1}{\lambda _{t}}\frac{d\lambda _{t}}{dt}=\frac{d}{dt}\ln
(1/\lambda _{t}).
\end{equation}

From the previous analysis we conclude that \textit{any non-Markovian
depolarizing evolution always fulfill the classicality constraints} defined
by Eqs.~(\ref{JoroClassPropa}) and~(\ref{ProbaMap}). In non-operational
approaches~\cite{BreuerReview,plenioReview} the presence of memory effects
can also be detected, for example, from the existence of time-intervals
where $\gamma _{t}<0$~\cite{canonicalCresser}. On the other hand, we remark
that the time-evolution~(\ref{RhoSEvolution}) does not cover the case in
which the dynamics is defined by a time-independent convex combination of
depolarizing maps [Eq.~(\ref{convex})], each one with a different unitary
dynamics. This case can only be worked out explicitly after specifying the
corresponding Hamiltonians (see Ref.~\cite{LindbladRate}).

\subsection{Operational approach}

In operational approaches the definition of classicality involves the
bipartite $s$-$e$ propagator. Similarly to Ref.~\cite{DNI}, we restrict the
present studies to bipartite propagators that fulfill a semigroup property, $%
\mathcal{G}_{t+\tau ,0}=\mathcal{G}_{t+\tau ,t}\mathcal{G}_{t,0}.$
Furthermore, their structures remain the same between measurement processes.
Hence, under these conditions, the bipartite $s$-$e$ state $\rho _{t}^{se}$
must evolves as%
\begin{equation}
\frac{d\rho _{t}^{se}}{dt}=\mathcal{L}[\rho _{t}^{se}],\ \ \ \ \ \ \ \ \rho
_{t}^{se}=\mathcal{G}_{t,0}[\rho _{0}^{se}],  \label{BipartitePropa}
\end{equation}%
where $\mathcal{G}_{t,0}=\exp (t\mathcal{L}).$ For simplicity, and without
loss of generality, here the superoperator $\mathcal{L}$ does not depend on
time. It could correspond to a (microscopic) unitary description of the $s$-$%
e$ coupling. Alternatively, it could correspond to a bipartite Lindblad
equation. In this case, the $s$-$e$ coupling is not unitary.

Considering an arbitrary time-generator $\mathcal{L},$ the classicality
condition Eq.~(\ref{Fixed}) could be satisfied for some particular dephasing
maps $\Delta .$ Nevertheless, a general characterization of this situation
is hard to obtain. A manageable simpler problem is to consider under which
conditions the bipartite $s$-$e$ dynamics does not generate any quantum
discord in the basis associated to $\Delta $ [condition that in turn
guarantees the fulfilment of Eq.~(\ref{Fixed})]. This problem was partially
studied in (bipartite) unitary \cite{poland} and (bipartite) Lindblad
dynamics~\cite{cero,almost}.

A general characterization could be obtained by imposing the validity of the
classicality condition Eq.~(\ref{Fixed}) for (fixed but) arbitrary dephasing
maps $\Delta .$ Nevertheless, this situation emerges as a particular case of
superclassicality (see examples below). On the other hand, we notice that
while the definition of classicality with a (unique) fixed basis $\Delta $
makes complete sense, the fulfillment of its definition is unstable with
respect to the system unitary evolution [consider for example the
depolarizing propagator Eq.~(\ref{RhoSistema}) with $U_{t}=0$ and $U_{t}\neq
0].$

\subsection{Superclassical open system evolutions}

In contrast to the previous approach, finding the conditions under which
superclassicality is valid becomes a simpler problem. This feature is mainly
due to the arbitrariness of the first and last measurement processes in the
definition~(\ref{DeltaSuper}). On the other hand, notice that here any extra
unitary system contribution, due to the definition of the intermediate
measurement [Eq.~(\ref{YtMeaurement})], does not affect the classicality
condition.

The bipartite $s$-$e$ propagator, even between measurement processes, is
defined by Eq.~(\ref{BipartitePropa}). As demonstrated in Ref.~\cite{DNI}
DNI cannot be fulfilled when the $s$-$e$ interaction is unitary or described
through stochastic Hamiltonians. Below, we determine the possible
non-unitary $s$-$e$ couplings. Depending on the production of quantum
discord different solutions emerge.

\subsubsection{Superclassicality without discord generation}

The bipartite propagator [Eq.~(\ref{BipartitePropa})] does not generate any
quantum discord when it assumes the structure%
\begin{equation}
\rho _{t}^{se}=\mathcal{G}_{t,0}[\rho _{0}\otimes \sigma
_{0}]=\sum_{c}|c_{t}\rangle \langle c_{t}|\otimes \sigma _{t}^{c}.
\label{NullDiscordPropa}
\end{equation}%
Here, $\{\sigma _{t}^{c}\}$ are non-normalized environment states while $%
\{|c_{t}\rangle \}$ is the basis where the system state is a diagonal matrix
at time $t,$ $\rho _{t}=\mathrm{Tr}_{e}(\mathcal{G}_{t,0}[\rho _{0}\otimes
\sigma _{0}])=\sum_{c}p_{t}^{c}|c_{t}\rangle \langle c_{t}|,$ where $%
p_{t}^{c}=\mathrm{Tr}_{e}(\sigma _{t}^{c}).$ Under the mapping $\rho
_{0}\rightarrow \rho _{X}=\bigtriangleup _{X}[\rho _{0}],$ it is simple to
check that the DNI property~(\ref{DeltaSuper}), and consequently the
Kolmogorov condition~(\ref{SuperCondition}), are identically fulfilled.
Nevertheless, this condition must be valid for arbitrary $X$-measurements
and initial conditions $\rho _{0}.$ Thus, superclassicality is fulfilled by
bipartite dynamics that \textit{does\ not generate any quantum discord
independently of which is the initial system state}. This property also
supports their superclassical denomination.

In order to find the time-evolutions that fulfill the previous condition, we
notice that the basis where the system state is diagonal at time $t$ and at
the initial time $t=0,$ $\rho _{0}=\sum_{c}p_{0}^{c}|c_{0}\rangle \langle
c_{0}|,$ can always be related by a unitary transformation%
\begin{equation}
|c_{t}\rangle =U_{t}|c_{0}\rangle .
\end{equation}%
Thus, Eq.~(\ref{NullDiscordPropa}) can always be rewritten as%
\begin{equation}
\rho _{t}^{se}=\sum_{c}U_{t}|c_{0}\rangle \langle c_{0}|U_{t}^{\dagger
}\otimes \sigma _{t}^{c}.
\end{equation}%
The system state, $\rho _{t}=\mathrm{Tr}_{e}(\rho _{t}^{se}),$ and the
environment state, $\sigma _{t}=\mathrm{Tr}_{s}(\rho _{t}^{se}),$
consequently read%
\begin{equation}
\rho _{t}=\sum_{c}U_{t}|c_{0}\rangle \langle c_{0}|U_{t}^{\dagger }\mathrm{Tr%
}_{e}(\sigma _{t}^{c}),\ \ \ \ \ \ \sigma _{t}=\sum_{c}\sigma _{t}^{c}.
\label{RhoSandE}
\end{equation}%
In general, the unitary transformation $U_{t}$ could depends on the $s$-$e$
initial state. Nevertheless, given that $\rho _{t}^{se}$ must obeys the
underlying time-evolution~(\ref{BipartitePropa}) and given that Eq.~(\ref%
{RhoSandE}) is valid for arbitrary system initial conditions, $U_{t}$ is
taken independent of $\rho _{0}.$ Consequently, based on the results of
Appendix~\ref{UniVsDe}, the map $\rho _{0}\rightarrow \rho _{t}$ can be read
as a \textit{composition of a unitary and a depolarizing map}. In fact, in
Eq.~(\ref{RhoSandE}) the unitary transformation changes the basis $%
|c_{0}\rangle \rightarrow U_{t}|c_{0}\rangle $ while the depolarizing map
changes the eigenvalues as $p_{0}^{c}\rightarrow p_{t}^{c}=\mathrm{Tr}%
_{e}(\sigma _{t}^{c}).$

In order to fulfill the previous property, based on Eq.~(\ref{DepolMap}), we
write the bipartite state [Eq.~(\ref{NullDiscordPropa})] as%
\begin{equation}
\rho _{t}^{se}=U_{t}\Big{\{}\rho _{0}\otimes \mathcal{E}_{t}[\sigma _{0}]+%
\frac{\mathrm{I}_{d}}{d}\otimes \mathcal{\bar{E}}_{t}[\sigma _{0}]\Big{\}}%
U_{t}^{\dagger }.  \label{BipartitePropal}
\end{equation}%
Here, $\mathcal{E}_{t}$ and $\mathcal{\bar{E}}_{t}$ are two completely
positive transformations acting on the initial environment state, with
initial conditions $\mathcal{E}_{0}[\sigma _{0}]=\sigma _{0}$ and $\mathcal{%
\bar{E}}_{0}[\sigma _{0}]=0.$ Normalization of the bipartite state $[\mathrm{%
Tr}_{se}(\rho _{t}^{se})=1]$ imposes the constraints%
\begin{equation}
\lambda _{t}=\mathrm{Tr}_{e}(\mathcal{E}_{t}[\sigma _{0}]),\ \ \ \ \ \ \ \ \
1-\lambda _{t}=\mathrm{Tr}_{e}(\mathcal{\bar{E}}_{t}[\sigma _{0}]).
\end{equation}%
Consistently, the null quantum discord condition Eq.~(\ref{NullDiscordPropa}%
) is fulfilled with%
\begin{equation}
|c_{t}\rangle =e^{-itH_{s}}|c_{0}\rangle ,\ \ \ \ \sigma _{t}^{c}=p_{0}^{c}%
\mathcal{E}_{t}[\sigma _{0}]+\mathcal{\bar{E}}_{t}[\sigma _{0}]/d,
\label{Base}
\end{equation}%
where we assumed that the system Hamiltonian does not depend on time, $%
U_{t}=e^{-itH_{s}}.$ Furthermore, it is simple to confirm that the system
state $[\rho _{t}=\mathrm{Tr}_{e}(\rho _{t}^{se})]$ assume the depolarizing
map structure Eq.~(\ref{RhoSistema}). Hence, its time evolution also obeys
Eq.~(\ref{RhoSEvolution}).

In order to find the underlying $s$-$e$ evolution [Eq.~(\ref{BipartitePropa}%
)] consistent with the propagator~(\ref{BipartitePropal}) we notice that the
environment state $\sigma _{t}=\mathcal{E}_{t}[\sigma _{0}]+\mathcal{\bar{E}}%
_{t}[\sigma _{0}]$ does not depend on the system degrees of freedom. We
assume that it obeys an arbitrary Lindblad evolution%
\begin{equation}
\frac{d\sigma _{t}}{dt}=\mathcal{L}_{e}[\sigma _{t}]+\sum_{\alpha }\gamma
_{\alpha }(B_{\alpha }\sigma _{t}B_{\alpha }^{\dagger }-\frac{1}{2}%
\{B_{\alpha }^{\dagger }B_{\alpha },\sigma _{t}\}_{+}),  \label{RhoEnviLin}
\end{equation}%
where $\{a,b\}_{+}=ab+ba$ is an anticommutator operation. $\mathcal{L}_{e}$
is an arbitrary extra (unitary or non-unitary) contribution. On the other
hand, in Eq.~(\ref{BipartitePropal}) $\mathcal{E}_{t}[\sigma _{0}]$ and $%
\mathcal{\bar{E}}_{t}[\sigma _{0}]$ can be read as the conditional
environment states given that the system remains in the initial condition or
changes to the maximal mixed state respectively. Thus, we propose the
bipartite evolution%
\begin{eqnarray}
\frac{d\rho _{t}^{se}}{dt} &=&-i[H_{s},\rho _{t}^{se}]+\mathcal{L}_{e}[\rho
_{t}^{se}]+\sum_{\alpha }\gamma _{\alpha }\Big{(}B_{\alpha }\mathcal{D}%
_{w_{\alpha }}[\rho _{t}^{se}]B_{\alpha }^{\dagger }  \notag \\
&&-\frac{1}{2}\Big{\{}B_{\alpha }^{\dagger }B_{\alpha },\rho _{t}^{se}%
\Big{\}}_{+}\Big{)},  \label{LindbladGen}
\end{eqnarray}%
where\ $\mathcal{D}_{w_{\alpha }}$ are depolarizing maps [Eq.~(\ref{DepolMap}%
)], each one with (arbitrary) weight $w_{\alpha }.$ This is the main result
of this section. It defines an underlying dissipative $s$-$e$ interaction
consistent with superclassicality and without discord generation. It can be
read as a collisional-like dynamics where the superoperators $\{\mathcal{D}%
_{\lambda _{\alpha }}\}$ are applied over the system whenever the
environment suffers a transition induced by the operators $\{B_{\alpha }\}.$

From Eq.~(\ref{LindbladGen}) it follows that $\sigma _{t}=\mathrm{Tr}%
_{s}(\rho _{t}^{se})$ obeys the evolution (\ref{RhoEnviLin}). On the other
hand, introducing Eq.~(\ref{BipartitePropal}) into Eq.~(\ref{LindbladGen}),
denoting $\eta _{t}\equiv \mathcal{E}_{t}[\sigma _{0}],\ \bar{\eta}%
_{t}\equiv \mathcal{\bar{E}}_{t}[\sigma _{0}],$ we get 
\begin{subequations}
\label{Auxiliares}
\begin{eqnarray}
\frac{d\eta _{t}}{dt} &=&\mathcal{L}_{e}[\eta _{t}]+\sum_{\alpha }\gamma
_{\alpha }w_{\alpha }(B_{\alpha }\eta _{t}B_{\alpha }^{\dagger }-\frac{1}{2}%
\{B_{\alpha }^{\dagger }B_{\alpha },\eta _{t}\}_{+})  \notag \\
&&-\sum_{\alpha }\gamma _{\alpha }(1-w_{\alpha })\frac{1}{2}\{B_{\alpha
}^{\dagger }B_{\alpha },\eta _{t}\}_{+}.
\end{eqnarray}%
Similarly%
\begin{eqnarray}
\frac{d\bar{\eta}_{t}}{dt} &=&\mathcal{L}_{e}[\bar{\eta}_{t}]+\sum_{\alpha
}\gamma _{\alpha }(B_{\alpha }\bar{\eta}_{t}B_{\alpha }^{\dagger }-\frac{1}{2%
}\{B_{\alpha }^{\dagger }B_{\alpha },\bar{\eta}_{t}\}_{+})  \notag \\
&&+\sum_{\alpha }\gamma _{\alpha }(1-w_{\alpha })B_{\alpha }\eta
_{t}B_{\alpha }^{\dagger }.
\end{eqnarray}%
It is simple to realize that these evolutions have the structure of a
generalized Lindblad equation~\cite{LindbladRate}. Thus, their evolution is
completely positive, which supports the vanishing discord solution defined
by Eqs.~(\ref{BipartitePropal}) and~(\ref{Base}). While the previous
derivation was performed starting in a totally uncorrelated $s$-$e$ state,
it is possible to prove that the evolution~(\ref{LindbladGen}) transforms
any initial state with null discord in a state with the same property.

\subsubsection{Superclassicality with discord generation}

Superclassicality can be satisfied even when the bipartite $s$-$e$
interaction leads to quantum discord [the solution structure~(\ref%
{NullDiscordPropa}) is not satisfied]. Here, we find which underlying $s$-$e$
interactions could lead to this situation.

The starting point is to note that the extra\ non-classical correlations
generated by the $s$-$e$ coupling must not affect the system dynamics~\cite%
{DNI}. Thus, the system propagator must be again the composition of a
unitary and a depolarizing map. Taking into account Eq.~(\ref%
{BipartitePropal}) we propose the propagator structure $[\rho _{t}^{se}=%
\mathcal{G}_{t,0}[\rho _{0}^{se}]]$%
\end{subequations}
\begin{equation}
\rho _{t}^{se}=U_{t}\Big{\{}\rho _{0}\otimes \mathcal{E}_{t}[\sigma
_{0}]+\sum_{k}W_{k}\rho _{0}W_{k}^{\dagger }\otimes \mathcal{E}%
_{t}^{k}[\sigma _{0}]\Big{\}}U_{t}^{\dagger },  \label{PropaDiscoGen}
\end{equation}%
where the Weyl-Heisenberg operators $\{W_{k}\}$ lead to maximal mixed state
[Eq.~(\ref{HWeyl})]. Similarly, $\mathcal{E}_{t}$ and $\{\mathcal{E}%
_{t}^{k}\}$ are completely positive transformations acting on the initial
environment state. It is simple to realize that here in general quantum
discord does not vanish.

Normalization of the bipartite state leads to the conditions%
\begin{equation}
\lambda _{t}=\mathrm{Tr}_{e}(\mathcal{E}_{t}[\sigma _{0}]),\ \ \ \ \ \ \ \ \
1-\lambda _{t}=\mathrm{Tr}_{e}(\mathcal{E}_{t}^{k}[\sigma _{0}])\ \forall k.
\label{PolarCondicion}
\end{equation}%
Here, the right equalities guaranty that the system propagator is again
given by Eq.~(\ref{RhoSistema}). In Appendix~\ref{OutcomeProbabil}, based on
the propagator~(\ref{PropaDiscoGen}), we calculate the outcome probabilities 
$P_{2}(z,x)$ and $P_{3}(z,y,x).$ After imposing the DNI conditions~(\ref%
{SuperCondition}) it follows that the environment propagators, added to the
constraints~(\ref{PolarCondicion}), must to fulfill that $\mathrm{Tr}_{e}(%
\mathcal{E}_{\tau }\mathcal{E}_{t}^{k}[\sigma _{0}]),$ $\mathrm{Tr}_{e}(%
\mathcal{E}_{\tau }^{k}\mathcal{E}_{t}[\sigma _{0}]),$ and $\mathrm{Tr}_{e}(%
\mathcal{E}_{\tau }^{k}\mathcal{E}_{t}^{j}[\sigma _{0}])$ do not depend on
indexes $k$ and $j.$ Furthermore, the relation%
\begin{eqnarray}
1-\mathrm{Tr}_{e}(\mathcal{E}_{t+\tau }[\sigma _{0}]) &=&\mathrm{Tr}_{e}(%
\mathcal{E}_{\tau }\mathcal{E}_{t}^{k}[\sigma _{0}])+\mathrm{Tr}_{e}(%
\mathcal{E}_{\tau }^{k}\mathcal{E}_{t}[\sigma _{0}])  \notag \\
&&+\mathrm{Tr}_{e}(\mathcal{E}_{\tau }^{k}\mathcal{E}_{t}^{j}[\sigma _{0}]),
\label{ConditionalDiscoPresente}
\end{eqnarray}%
must be fulfilled.

The conditions~(\ref{PolarCondicion}), which guaranty a depolarizing system
dynamics, and the conditions~(\ref{ConditionalDiscoPresente}), which
guaranty DNI even in presence of quantum discord, impose a severe constraint
on the possible $s$-$e$ bipartite propagator [Eq.~(\ref{PropaDiscoGen})].
These constraints cannot be checked without having an explicit expression
for the environment propagators $\mathcal{E}_{t}$ and $\{\mathcal{E}%
_{t}^{k}\}.$ Therefore, in this case, it is not possible to derive a general
underlying bipartite $s$-$e$ evolution consistent with these constraints.
Nevertheless, taking the structure of Eqs.~(\ref{BipartitePropal}) and~(\ref%
{LindbladGen}), here we conjecture the bipartite evolution%
\begin{equation}
\frac{d\rho _{t}^{se}}{dt}=\sum_{\alpha ,k}\gamma _{\alpha }(B_{\alpha }^{k}%
\mathcal{D}_{w_{\alpha }}^{k}[\rho _{t}^{se}]B_{\alpha }^{k\dagger }-\frac{1%
}{2}\{B_{\alpha }^{k\dagger }B_{\alpha }^{k},\rho _{t}^{se}\}_{+}),
\label{DiscoGenera}
\end{equation}%
where the \textquotedblleft partial depolarizing maps\textquotedblright\ are 
$\mathcal{D}_{w_{\alpha }}^{k}[\rho ]\equiv w_{\alpha }\rho \delta
_{k0}+(1-w_{\alpha })W_{k}\rho W_{k}^{\dagger }.$ This map is applied over
the system whenever the environment suffers a transition induced by the
operators $B_{\alpha }^{k}.$ In Eq.~(\ref{DiscoGenera}), a system
Hamiltonian $H_{s}$ is omitted because it cannot commutates with all
operators $\{B_{\alpha }^{k}\},$ which lead to an inconsistence with the
constraints~(\ref{PolarCondicion}) and~(\ref{ConditionalDiscoPresente}).
When $B_{\alpha }^{k}=B_{\alpha }$ we recover the evolution~(\ref%
{LindbladGen}), which does not generate quantum discord. The structure~(\ref%
{DiscoGenera}) is supported by the examples studied below.

\section{Examples \label{SecExamples}}

In this section we provide a set of specific examples that lighten the
different definitions of classicality. The system is a qubit. Its density
matrix reads [see Eq.~(\ref{RhoSistema})],%
\begin{equation}
\rho _{t}=w(t)\rho _{0}+\frac{[1-w(t)]}{3}\mathcal{S}[\rho _{0}],
\label{PropaTLS}
\end{equation}%
where $w(t)$ is a time-dependent positive weight. For simplicity, the system
Hamiltonian vanishes identically. On the other hand, the superoperator $%
\mathcal{S}$ is%
\begin{equation}
\mathcal{S}[\bullet ]\equiv \sum_{k=1,2,3}\mathcal{S}_{k}[\bullet
]=\sum_{k=1,2,3}\sigma _{k}\bullet \sigma _{k}.  \label{Sk}
\end{equation}%
Correspondingly with Eq.~(\ref{HWeyl}), it can be rewritten as $\mathcal{S}%
[\rho ]=(2\mathrm{I}_{s}-\rho ).$ With $\sigma _{k}$ $(k=1,2,3)$ we denote
the (system) Pauli matrices. The propagator~(\ref{PropaTLS}) has associated
the master equation%
\begin{equation}
\frac{d\rho _{t}}{dt}=\gamma _{t}(\mathcal{S}[\rho _{t}]-\rho _{t}),
\label{GamaTLS}
\end{equation}%
where the time-dependent rate here explicitly reads $\gamma
_{t}=[1-4w(t)]^{-1}(d/dt)w(t).$

As the system dynamics is defined by a depolarizing map, the non-operational
definition of classicality is always satisfied [Eqs.~(\ref{JoroClassPropa})
and~(\ref{ProbaMap})]. Nevertheless, the operational definition with fixed
basis [Eq.~(\ref{Fixed})] and superclassicality [Eq.~(\ref{DeltaSuper})] are
not always satisfied. Departure from this last condition is measured with
the distance $I(t,\tau )$ [Eq.~(\ref{IMeasure})]. The three (projective)
measurements [performed at times $0,$ $t,$ and $t+\tau ,$ which have
associated the joint probability $P_{3}(z,y,x)]$ are specified through their
polar angles in the Bloch sphere~\cite{nielsen}, $\{\theta _{M},\phi _{M}\},$
$M=X,Y,Z.$ For simplicity, in all cases, we take $\phi _{M}=0.$ The possible
values of each observable are $x=\pm 1,\ y=\pm 1,\ $and $z=\pm 1.$ Given
that the depolarizing dynamics does not change the basis where the system
state is a diagonal matrix, the intermediate $Y$-measurement that commutates
with the pre-measurement state [Eq.~(\ref{YtMeaurement})] is defined by
taking $\theta _{Y}=\theta _{X},$ while $\theta _{X}$ and $\theta _{Z}$
remain as arbitrary angles. On the other hand, in order to establishing the
presence of memory effects we use the conditional past-future correlation~(%
\ref{CPFDefinition}). Non-Markovianity translates to the existence of at
least a set of (unrestricted) angles $\theta _{X},$ $\theta _{Y},$ $\theta
_{Z},$ such that $C_{pf}(t,\tau )\neq 0.$ For each of the following models,
the outcome probabilities can be calculated in an exact analytical way from
standard quantum measurement theory (see also Appendix~\ref{OutcomeProbabil}%
).

\subsection{Depolarizing superclassical dynamics}

Over the basis of the results defined by Eqs.~(\ref{LindbladGen}) and~(\ref%
{DiscoGenera}) we formulate different superclassical dynamics with and
without quantum discord respectively.

\subsubsection{Superclassical dynamics without discord generation}

The bipartite $s$-$e$ state evolves as%
\begin{equation}
\frac{d\rho _{t}^{se}}{dt}=\gamma (B\mathcal{S}[\rho _{t}^{se}]B^{\dagger }-%
\frac{1}{2}\{B^{\dagger }B,\rho _{t}^{se}\}_{+}).  \label{DecayDNull}
\end{equation}%
The environment is also a two-level system, whose states are denoted as $%
\{|0\rangle ,|1\rangle \}.$ The environment operator is $B=|1\rangle \langle
0|.$

The bipartite evolution can be read straightforwardly as follows. The
operator $B$ induces the environment transition $|0\rangle \overset{\mathcal{%
S}}{\rightarrow }|1\rangle ,$ which simultaneously implies that the
superoperator $\mathcal{S}$ is applied over the system. The transitions
occur at random times. The rate $\gamma $ sets their exponential statistics.

Considering the initial bipartite state%
\begin{equation}
\rho _{0}^{se}=\rho _{0}\otimes \sigma _{0}=\rho _{0}\otimes |0\rangle
\langle 0|,
\end{equation}%
the solution of Eq.~(\ref{DecayDNull}) is%
\begin{equation}
\rho _{t}^{se}=w(t)\rho _{0}\otimes |0\rangle \langle 0|+\frac{1-w(t)}{3}%
\Big{(}\!\sum_{k=1,2,3}\mathcal{S}_{k}[\rho _{0}]\Big{)}\otimes |1\rangle
\langle 1|,  \label{ExampleLin}
\end{equation}%
where the weight is $w(t)=\exp (-\gamma t).$ Given that $\rho _{0}$ and $%
\mathcal{S}[\rho _{0}]$ are always diagonal in the same basis, it is simple
to check that quantum discord is not generated even when considering
arbitrary system initial conditions $\rho _{0}.$ This property guarantees
the superclassical character of the dynamics.

Using the propagator~(\ref{ExampleLin}), the measure of deviations with
respect to DNI [Eq.~(\ref{IMeasure})] explicitly reads%
\begin{equation}
I(t,\tau )=\frac{1}{3}(4-e^{-\gamma (t+\tau )})|\sin (\theta _{Z}-\theta
_{Y})\sin (\theta _{Y}-\theta _{X})|,  \label{I2}
\end{equation}%
expression valid for arbitrary $\rho _{0}.$ Consistently, when $\theta
_{Y}=\theta _{X},$ the intermediate measurement does not perturb the
evolution, $I(t,\tau )=0.$ In fact, this is the unique election that
guaranty DNI when considering \textit{arbitrary} angles $\theta _{X}$ and $%
\theta _{Z}.$ These properties confirm the superclassical character of the
system dynamics [Eq.~(\ref{SuperCondition})]. On the other hand, notice that
in this example superclassicality implies that the classicality definition~(%
\ref{Fixed}) is fulfilled in any measurement (fixed) basis (defined by the
angle $\theta =\theta _{X}=\theta _{Y}=\theta _{Z}).$

The validity of DNI occurs in presence of memory effects. For example, the
rate in Eq.~(\ref{GamaTLS}) is characterized by \textquotedblleft
non-Markovian\textquotedblright\ divergent behaviors $\gamma (t)=\gamma
/(4-e^{\gamma t}).$ On the other hand, assuming an initial system state $%
\rho _{0}$ such that $P_{1}(x)=1/2,$ the CPF correlation [Eq.~(\ref%
{CPFDefinition})] reads%
\begin{eqnarray}
C_{pf}(t,\tau ) &=&-\frac{16}{9}e^{-\gamma t}(1-e^{-\gamma t})(1-e^{-\gamma
\tau })  \label{CPF2} \\
&&\times \cos (\theta _{Z}-\theta _{Y})\cos (\theta _{Y}-\theta _{X}). 
\notag
\end{eqnarray}%
The property $C_{pf}(t,\tau )\neq 0$ indicates the presence of memory
effects. In this example, when $\theta _{Y}=\theta _{X}$ (the DNI basis) the
memory effects, measured by $C_{pf}(t,\tau ),$ are maximal.

\subsubsection{Superclassical dynamics with discord generation}

In this example, the environment Hilbert space is four dimensional with a
natural basis of states $\{|0\rangle ,|1\rangle ,|2\rangle ,|3\rangle \}.$
Consistently with Eq.~(\ref{DiscoGenera}), the $s$-$e$ dynamics reads%
\begin{equation}
\frac{d\rho _{t}^{se}}{dt}=\frac{\gamma }{3}\sum_{k=1,2,3}B_{k}\mathcal{S}%
_{k}[\rho _{t}^{se}]B_{k}^{\dagger }-\frac{1}{2}\{B_{k}^{\dagger }B_{k},\rho
_{t}^{se}\}_{+}.  \label{ConDisco4}
\end{equation}%
Here, the bath operators are $B_{k}\equiv |k\rangle \langle 0|,$ $k=1,2,3.$
The partial depolarizing maps $\mathcal{S}_{k}$ are defined in Eq.~(\ref{Sk}%
). These maps are applied over the system whenever the environment suffer
the transition $|0\rangle \overset{\mathcal{S}_{k}}{\rightarrow }|k\rangle ,$
which is induced by the operator $B_{k}.$ Hence, the dynamics is similar to
that defined by Eq. (\ref{DecayDNull}) but here a different bath transition
is associated to each superoperator $\mathcal{S}_{k}.$

Taking the initial state%
\begin{equation}
\rho _{0}^{se}=\rho _{0}\otimes \sigma _{0}=\rho _{0}\otimes |0\rangle
\langle 0|,  \label{CIN4}
\end{equation}%
the solution of Eq.~(\ref{ConDisco4}) reads%
\begin{equation}
\rho _{t}^{se}=w(t)\rho _{0}\otimes |0\rangle \langle 0|+\frac{1-w(t)}{3}%
\Big{(}\!\sum_{k=1,2,3}\mathcal{S}_{k}[\rho _{0}]\otimes |k\rangle \langle k|%
\Big{)},  \label{BipaConDisco}
\end{equation}%
where the weight is $w(t)=\exp (-\gamma t).$ From this expression it is
possible to check that the conditions~(\ref{PolarCondicion}) are satisfied.

The system state $\rho _{t}=\mathrm{Tr}_{e}(\rho _{t}^{se})$ is exactly the
same than in Eq.~(\ref{ExampleLin}). Nevertheless, in contrast, here quantum
discord is generated after the initial time.\ In spite of this property,
performing an explicit calculus based on the bipartite propagator~(\ref%
{BipaConDisco}), it is possible to demonstrate that the distance $I(t,\tau )$
and the CPF correlation are exactly the same than in the previous case,
Eqs.~(\ref{I2}) and~(\ref{CPF2}) respectively. This equivalence demonstrates
that the dynamics is superclassical but with discord generation.

\subsection{Depolarizing dynamics that are not superclassical}

In general, depolarizing system dynamics do not lead to superclassicality.
Here, we study different examples.

\subsubsection{Incoherent reservoir dynamics}

Taking into account the general result defined by Eq.~(\ref{LindbladGen}),
the superclassical example without quantum discord generation [Eq.~(\ref%
{DecayDNull})] can be generalized as%
\begin{eqnarray}
\frac{d\rho _{t}^{se}}{dt} &=&\gamma (B\mathcal{S}[\rho _{t}^{se}]B^{\dagger
}-\frac{1}{2}\{B^{\dagger }B,\rho _{t}^{se}\}_{+})  \label{General} \\
&&+\phi (B^{\dagger }\mathcal{S}[\rho _{t}^{se}]B-\frac{1}{2}\{BB^{\dagger
},\rho _{t}^{se}\}_{+}).  \notag
\end{eqnarray}%
Here the two-level environment dynamics is defined by the transitions $%
|0\rangle \overset{\mathcal{S}}{\rightarrow }|1\rangle $ and $|1\rangle 
\overset{\mathcal{S}}{\rightarrow }|0\rangle $ with rates $\gamma $ and $%
\phi $ respectively. In each transition it is applied the superoperator $%
\mathcal{S}$. This dynamics, even including an arbitrary system Hamiltonian,
is superclassical without discord generation. Nevertheless, the addition of
an arbitrary system Hamiltonian makes impossible to fulfill classicality
definition with fixed basis.

In contrast to the previous example [Eq.~(\ref{General})], the general
structure Eq.~(\ref{DiscoGenera}) is strongly constrained by the conditions~(%
\ref{ConditionalDiscoPresente}). For example, we generalize the
superclassical dynamics~(\ref{ConDisco4}) as%
\begin{eqnarray}
\frac{d\rho _{t}^{se}}{dt} &=&\frac{\gamma }{3}\!\!\!\sum_{k=1,2,3}\!\!%
\!B_{k}\mathcal{S}_{k}[\rho _{t}^{se}]B_{k}^{\dagger }-\frac{1}{2}%
\{B_{k}^{\dagger }B_{k},\rho _{t}^{se}\}_{+} \\
&&+\phi \!\!\sum_{k=1,2,3}(B_{k}^{\dagger }\mathcal{S}_{k}[\rho
_{t}^{se}]B_{k}-\frac{1}{2}\{B_{k}B_{k}^{\dagger },\rho _{t}^{se}\}_{+}. 
\notag
\end{eqnarray}%
The rates $\gamma $\ and $\phi $\ are associated to the transitions $%
|0\rangle \overset{\mathcal{S}_{k}}{\rightarrow }|k\rangle $ and $|k\rangle 
\overset{\mathcal{S}_{k}}{\rightarrow }|0\rangle $\ respectively. Taking the
initial condition~(\ref{CIN4}), the system depolarizing dynamics is defined
with the weight $w(t)=[\phi +\gamma e^{-t(\gamma +\phi )}]/(\gamma +\phi ).$
Nevertheless, in this case the constraints~(\ref{ConditionalDiscoPresente})
are not fulfilled. Thus, the superclassical property is lost. In fact, for
simplicity, defining the three measurements with $\theta _{Z}=\theta
_{Y}=\theta _{X},$ departures from DNI [Eq.~(\ref{IMeasure})] are measured by%
\begin{equation}
I(t,\tau )=|g(t,\tau )|\sin ^{2}(2\theta _{X}),
\end{equation}%
where $g(t,\tau )$ is a real function that depends parametrically on $\gamma 
$ and $\phi .$ For $\phi =\gamma /3,$ it is $g(t,\tau )=(1/3)e^{-(\gamma
/3)\tau }(1-e^{-\gamma \tau })(1-e^{-(4/3)\gamma t}).$ Thus, DNI is not
fulfilled because, even when the intermediate measurement commutates with
the pre-measurement state, it is not possible to guaranty that $I(t,\tau )=0$
for arbitrary $\theta _{X}.$ On the other hand, the definition~(\ref{Fixed})
is only fulfilled when $\theta _{X}=0.$

The CPF correlation, under the same assumptions $(\theta _{Z}=\theta
_{Y}=\theta _{X}),$ $\phi =\gamma /3,$ jointly with $P_{1}(x)=1/2,$ is%
\begin{equation}
C_{pf}(t,\tau )=-(1/4)g(t,\tau )[1-\cos (4\theta _{X})+8e^{-(4/3)\gamma t}],
\end{equation}%
which explicitly demonstrate that the outcome statistics is non-Markovian $%
[C_{pf}(t,\tau )\neq 0].$

\subsubsection{Unitary system-environment dynamics}

Now we consider a unitary $s$-$e$ evolution%
\begin{equation}
\frac{d\rho _{t}^{se}}{dt}=-i[H_{se},\rho _{t}^{se}],
\label{UnitaryEvolution}
\end{equation}%
where the Hamiltonian $H_{se}$ introduces their mutual interaction. For
simplicity we consider a two-level environmental system. The Hamiltonian is
taken as%
\begin{equation}
H_{se}=\frac{\Omega }{2}\sum_{k=1,2,3}\sigma _{k}\otimes \sigma _{k}.
\label{Hse_SingleBath}
\end{equation}%
With $\sigma _{k}$ we denotes the $k$-Pauli matrix $(k=1,2,3)$ in both the
system and environment Hilbert spaces. In order to obtain a depolarizing
system dynamic we take the bipartite initial condition%
\begin{equation}
\rho _{0}^{se}=\rho _{0}\otimes \sigma _{0}=\rho _{0}\otimes \mathrm{I}%
_{e}/2.
\end{equation}%
Thus, the environment is always initialized in the maximal mixed state $%
\mathrm{I}_{e}/2$ while $\rho _{0}$ is arbitrary.

The solution of Eq.~(\ref{UnitaryEvolution}) can be written as%
\begin{equation}
\rho _{t}^{se}=e^{-itH_{se}}\rho _{0}^{se}e^{+itH_{se}}.
\label{UnitarioBipaPropa}
\end{equation}%
The propagator for bipartite wave vectors is $e^{-itH_{se}}=(a_{t}+b_{t})%
\mathrm{I}_{se}-b_{t}\sum_{k=1,2,3}\sigma _{k}\otimes \sigma _{k},$ where $%
a_{t}=\exp (-i\Omega t/2)$ and $b_{t}=(i/2)\exp (+i\Omega t/2)\sin (\Omega
t).$ These expressions solve the bipartite state defined by~(\ref%
{UnitarioBipaPropa}). This dynamics does not assume the form~(\ref%
{BipartitePropal}) neither~(\ref{PropaDiscoGen}). In fact, in general the
interaction Hamiltonian generates both quantum discord and quantum
entanglement.

The depolarizing system dynamics [Eq.~(\ref{PropaTLS})] is defined by the
weight $w(t)=(1/4)[1+3\cos ^{2}(\Omega t)].$ On the other hand, departures
from DNI [Eq.~(\ref{IMeasure})], for arbitrary measurements defined by the
angles $\theta _{Z}\neq \theta _{Y}\neq \theta _{X},$ read%
\begin{eqnarray}
I(t) &=&\Big{|}\frac{1}{2}\cos (\theta _{Z}-\theta _{X})[\cos (2t\Omega
)+\cos (4t\Omega )]  \notag \\
&&-\cos (\theta _{Z}-2\theta _{Y}+\theta _{X})[\cos (\Omega t)]^{2}\Big{|}.
\label{DistanceSingleBath}
\end{eqnarray}%
For simplifying the expression, we have taken equal time intervals between
measurements, $\tau =t,$ where $I(t,t)\rightarrow I(t).$ This distance,
under the DNI\ measurement condition $\theta _{Y}=\theta _{X}$ does not
vanish. In fact, it reduces to%
\begin{equation}
I(t)=|\cos (\theta _{Z}-\theta _{X})|\sin ^{2}(2t\Omega ).
\end{equation}%
Notice that, it is not possible to guaranty $I(t)=0$ for arbitrary angles $%
\theta _{Z}$ and $\theta _{X}.$ Hence, DNI is not satisfied.

Under the same DNI condition $[\theta _{Y}=\theta _{X}],$ the CPF
correlation $(\tau =t)$ [Eq.~(\ref{CPFDefinition})] reads $%
[C_{pf}(t,t)\rightarrow C_{pf}(t)]$%
\begin{equation}
C_{pf}(t)=\cos (\theta _{Z}-\theta _{X})\sin ^{4}(\Omega t).
\end{equation}%
Thus, the dynamics develops memory effects and does not obey DNI. Using
similar calculus, these properties remain the same if one take into account
an environment consisting of more than one qubit.

\section{Summary and Conclusions}

In this paper, we characterized a class of superclassical non-Markovian open
quantum dynamics and, in addition, compared its definition with previous
notions of classicality in both non-operational~\cite{horoNoMark} and
operational approaches~\cite{plenioNoMarkov}.

The non-operational definition of classicality introduced in Ref.~\cite%
{horoNoMark} requires that the system propagator can be written as the
composition of a classical and a unitary map [Eq.~(\ref{JoroClassPropa})].
On the other hand, the operational definition of classicality introduced in
Ref.~\cite{plenioNoMarkov} requires that, in a given fixed measurement
basis, an intermediate measurement processes does not affect the system
dynamics [Eq.~(\ref{Fixed})]. Superclassicality is also defined in an
operational way, being a much restrictive condition. Indeed, in this notion
of classicality, an intermediate measurement process commuting with the
pre-measurement state must be noninvasive even when arbitrary previous and
posterior measurement processes are taken into account [Eqs.~(\ref%
{DeltaSuper}) and~(\ref{YtMeaurement})]. Therefore, Kolmogorov consistence
condition is satisfied under the same conditions [Eq.~(\ref{SuperCondition}%
)].

After providing a clear measurement-based definition of superclassicality,
we established the existence of specific $s$-$e$ models that leads to this
feature. Considering local-in-time bipartite Lindblad evolutions\ [Eq.~(\ref%
{BipartitePropa})] superclassicality is satisfied when the (non-unitary) $s$-%
$e$ coupling does not generate quantum discord for arbitrary system
initializations [Eq.~(\ref{NullDiscordPropa})]. We find that this strong
constraint is satisfied by a class of collisional-type dynamics where a
depolarizing map is applied over the system every time the environment
undergoes a transition [Eq.~(\ref{LindbladGen})]. Superclassicality can be
satisfied even if there is discord generation. However, extra propagator
constraints, which in general are hard to fulfill [Eq.~(\ref%
{ConditionalDiscoPresente})], limit the possible corresponding
time-evolutions [Eq.~(\ref{DiscoGenera})].

With or without discord generation, superclassicality restrict the system
dynamics to a subclass of non-Markovian depolarizing time-evolutions. This
feature is inherited from the arbitrariness of the measurement processes in
their definition [Eq.~(\ref{DeltaSuper})]. It was shown that superclassical
dynamics are also classical under the non-operational definition~\cite%
{horoNoMark} and, disregarding any Hamiltonian system contribution, they
also fulfill the classicality definition~\cite{horoNoMark} in an arbitrary
measurement basis.

We characterized different examples that fulfill superclassicality.
Furthermore, in consistency with Ref.~\cite{DNI}, we specified different
non-Markovian depolarizing dynamics that are not superclassical (both for
unitary and non-unitary $s$-$e$ couplings). The developed results support
that, with the exception of superclassical dynamics, non-Markovian memory
effects lead to an intrinsic measurement invasiveness that can be detected
with a minimal operational scheme based on performing three consecutive
measurement processes over the system of interest.

\section*{Acknowledgments}

This paper was supported by Consejo Nacional de Investigaciones Cient\'{\i}%
ficas y T\'{e}cnicas (CONICET), Argentina.

\appendix

\section{Unitary vs. depolarizing maps\label{UniVsDe}}

Given an arbitrary system state $\rho $ it can always be written as%
\begin{equation}
\rho =\sum_{c}p_{c}|c\rangle \langle c|.  \label{RhoIn}
\end{equation}%
The set of states $\{|c\rangle \}$ define the basis where $\rho $ is\
diagonal matrix while $\{p_{c}\}$ are the corresponding eigenvalues. Here we
characterize the more general maps whose action solely consists in modifying
the eigenvalues \textit{or} the eigenvectors of the input state $\rho .$

A map that lefts \textit{unmodified the eigenvalues} of the arbitrary input
state $\rho $ [Eq.~(\ref{RhoIn})] leads to the transformation%
\begin{equation}
\mathcal{U}[\rho ]=\sum_{c}p_{c}|\tilde{c}\rangle \langle \tilde{c}|.
\label{Unitary}
\end{equation}%
Here, the set of states $\{|\tilde{c}\rangle \}$ is also an orthogonal
basis. Thus, it is always possible to relate both basis by means of a
unitary transformation, $|\tilde{c}\rangle =U|c\rangle ,$ with $UU^{\dag }=%
\mathrm{I}_{s}.$ Here, $\mathrm{I}_{s}$ is the identity matrix in the system
Hilbert space. Using this property it follows that $\mathcal{U}[\rho
]=\sum_{c}p_{c}U|c\rangle \langle c|U^{\dagger }=U\rho U^{\dagger },$ which
leads to%
\begin{equation}
\mathcal{U}[\rho ]=U\rho U^{\dagger }.  \label{UnitaryMap}
\end{equation}%
Consistently, the demanded condition is solely fulfilled by \textit{unitary
transformations}.

In contrast with Eq.~(\ref{Unitary}), a map that left \textit{unmodified the
eigenvectors} of $\rho $ reads%
\begin{equation}
\mathcal{D}[\rho ]=\sum_{c}\tilde{p}_{c}|c\rangle \langle c|,
\label{DEigenvalues}
\end{equation}%
where the transformed eigenvalues $(p_{c}\rightarrow \tilde{p}_{c})$ satisfy 
$\sum_{c}\tilde{p}_{c}=1.$ Eq.~(\ref{DEigenvalues}) must be satisfied for
arbitrary inputs states. For finding the more general structure $\mathcal{D}$
we apply a unitary transformation to the previous expression,%
\begin{equation}
U\mathcal{D}[\rho ]U^{\dagger }=\sum_{c}\tilde{p}_{c}U|c\rangle \langle
c|U^{\dagger }=\mathcal{D}[U\rho U^{\dagger }].
\end{equation}%
Consequently, $\mathcal{D}$ must to commutate with any unitary map $\mathcal{%
U}.$ Two possible solutions are find, that is, $\mathcal{D}[\rho ]\approx
\rho \ $and$\ \mathcal{D}[\rho ]\approx \mathrm{I}_{s}.$ Thus, demanding
trace preservation, it follows the solution%
\begin{equation}
\mathcal{D}_{\lambda }[\rho ]=\lambda \rho +(1-\lambda )\frac{\mathrm{I}_{s}%
}{d},  \label{DepolarizingApp}
\end{equation}%
where $0\leq \lambda \leq 1$ is an arbitrary parameter while $d$ is the\
dimension of the system Hilbert space. By using that $\sum_{c}|c\rangle
\langle c|=\mathrm{I}_{s},$ the renewed eigenvalues [Eq.~(\ref{DEigenvalues}%
)] read $\tilde{p}_{c}=\lambda p_{c}+(1-\lambda )/d.$

We notice that $\mathcal{D}_{\lambda }$ is a standard \textit{depolarizing
map}~\cite{nielsen}. In contrast to unitary transformations, it is the
unique map that lefts invariant the basis where the input state is a
diagonal matrix. On the other hand, notice that unitary and depolarizing
transformations always commutate, $\mathcal{D}_{\lambda }\mathcal{U}=%
\mathcal{UD}_{\lambda }.$

\section{Measurement outcome-probabilities\label{OutcomeProbabil}}

Here, we calculate the outcomes probabilities $P_{3}(z,y,x)$ [Eq.~(\ref%
{P3Conjunta})] assuming that the underlying $s$-$e$ propagator is given by
Eq.~(\ref{PropaDiscoGen}), that is, superclassicality with discord
generation. The case in which there is not discord generation, Eq.~(\ref%
{BipartitePropal}), can be worked out along similar lines.

At the initial stage, $t=0,$ the bipartite initial state is $\rho
_{0}^{se}=\rho _{0}\otimes \sigma _{0}.$ In order to simplify the notation,
Eq.~(\ref{PropaDiscoGen}) is rewritten as%
\begin{equation}
\rho _{t}^{se}=U_{t}\rho _{0}U_{t}^{\dagger }\otimes \mathcal{E}_{t}[\sigma
_{0}]+\sum_{k}\mathcal{S}_{k}^{t}[U_{t}\rho _{0}U_{t}^{\dagger }]\otimes 
\mathcal{E}_{t}^{k}[\sigma _{0}],  \label{Solution}
\end{equation}%
where we have defined the system superoperators%
\begin{equation}
\mathcal{S}_{k}^{t}[\rho ]\equiv (U_{t}W_{k}U_{t}^{\dagger })\rho
(U_{t}W_{k}^{\dagger }U_{t}^{\dagger }).
\end{equation}

Each measurement processes is defined by a set of positive operator-valued
measure $\{E_{m}\}$ and post-measurement states $\{\rho _{m}\},$ where $m$
denotes the possible outcomes in each stage, $m=x,y,z.$ Given that all
measurements are projective, $E_{m}$ and $\rho _{m}$ are equal and
correspond to the projectors associated to each (non-degenerate) eigenvalue
of the corresponding observable. From Bayes rule it is always possible to
write%
\begin{equation}
P_{3}(z,y,x)=P_{3}(z|y,x)P_{2}(y|x)P_{1}(x),  \label{JointNoMarkov}
\end{equation}%
where $P_{n}(b|a)$ denotes the conditional probability of $b$ given $a$
while the subindex $n$ indicates the number of performed measurement
processes. Using that $\rho _{0}^{se}=\rho _{0}\otimes \sigma _{0},$ the
probability for the first measurement outcomes is $P_{1}(x)=\mathrm{Tr}%
_{s}(E_{x}\rho _{0}),$\ while the post-measurement bipartite state is $\rho
_{x}^{se}=\rho _{x}\otimes \sigma _{0}.$ At time $t$ the bipartite state, $%
\rho _{x}^{se}\rightarrow \rho _{x}^{se}(t),$ from Eq.~(\ref{Solution}) is%
\begin{equation}
\rho _{x}^{se}(t)=\rho _{t}^{x}\otimes \mathcal{E}_{t}[\sigma _{0}]+\sum_{k}%
\mathcal{S}_{k}^{t}[\rho _{t}^{x}]\otimes \mathcal{E}_{t}^{k}[\sigma _{0}].
\end{equation}%
Here, we defined the time-dependent system state $\rho _{t}^{m}\equiv
U_{t}\rho _{m}U_{t}^{\dagger },$ where $\rho _{m}$ is a post-measurement
system state.

The conditional probability for the next measurement reads $P_{2}(y|x)=%
\mathrm{Tr}_{se}[E_{y}\rho _{x}^{se}(t)],$ which delivers%
\begin{equation}
P_{2}(y|x)=\mathrm{Tr}_{s}(E_{y}\rho _{t}^{x})\mathrm{Tr}_{e}(\mathcal{E}%
_{t}[\sigma _{0}])+\frac{1-\mathrm{Tr}_{e}(\mathcal{E}_{t}[\sigma _{0}])}{d}.
\label{P(y|x)}
\end{equation}%
Here, we used the condition~(\ref{PolarCondicion}) and the relation~(\ref%
{HWeyl}). After the measurement, the bipartite changes disruptively $\rho
_{x}^{se}(t)\rightarrow \rho _{yx}^{se}(t),$ where%
\begin{equation}
\rho _{yx}^{se}(t)=\rho _{y}\otimes \sigma _{t}^{yx}.
\end{equation}%
The environment post-measurement state is%
\begin{equation}
\sigma _{t}^{yx}=\frac{\mathrm{Tr}_{s}(E_{y}\rho _{t}^{x})\mathcal{E}%
_{t}[\sigma _{0}]+\sum_{k}\mathrm{Tr}_{s}(E_{y}\mathcal{S}_{k}^{t}[\rho
_{t}^{x}])\mathcal{E}_{t}^{k}[\sigma _{0}]}{P_{2}(y|x)}.
\label{RhoEnvPostYX}
\end{equation}

The conditional probability for the third measurement is $P_{3}(z|y,x)=%
\mathrm{Tr}_{se}[E_{z}\rho _{yx}^{se}(t+\tau )],$ where $\rho
_{yx}^{se}(t+\tau )$ is the bipartite state evolved during a time interval $%
\tau $ with the propagator~(\ref{Solution}) taking the initial condition $%
\rho _{yx}^{se}(t).$ It follows%
\begin{equation}
\rho _{yx}^{se}(t+\tau )=\rho _{\tau }^{y}\otimes \mathcal{E}_{\tau }[\sigma
_{t}^{yx}]+\sum_{k}\mathcal{S}_{k}^{\tau }[\rho _{\tau }^{y}]\otimes 
\mathcal{E}_{\tau }^{k}[\sigma _{t}^{yx}].
\end{equation}%
Therefore,%
\begin{eqnarray}
P_{3}(z|y,x) &=&\mathrm{Tr}_{s}(E_{z}\rho _{\tau }^{y})\mathrm{Tr}_{e}(%
\mathcal{E}_{\tau }[\sigma _{t}^{yx}])  \label{P(z|yx)} \\
&&+\sum_{k}\mathrm{Tr}_{s}(E_{z}\mathcal{S}_{k}^{\tau }[\rho _{\tau }^{y}])%
\mathrm{Tr}_{e}(\mathcal{E}_{\tau }^{k}[\sigma _{t}^{yx}]),  \notag
\end{eqnarray}%
where $\sigma _{t}^{yx}$ is given by Eq.~(\ref{RhoEnvPostYX}).

\subsection*{Superclassicality condition}

The DNI condition~(\ref{SuperCondition}) gives the superclassical character
of a given non-Markovian dynamics. Using Eq.~(\ref{JointNoMarkov}), in terms
of conditional probabilities it becomes%
\begin{equation}
P_{3}(z|x)\overset{DNI}{=}\sum_{y}P_{3}(z|y,x)P_{2}(y|x)=P_{2}(z|x),
\label{KolmogorovConditional}
\end{equation}%
where the intermediate measurement commutates with the pre-measurement state
[Eq.~(\ref{YtMeaurement})], $E_{y}\rightarrow E_{y}^{t}.$

Under the replacements $t\rightarrow t+\tau $ and $y\rightarrow x,$ from
Eq.~(\ref{P(y|x)}) it follows that%
\begin{equation}
P_{2}(z|x)=\mathrm{Tr}_{s}(E_{z}\rho _{t+\tau }^{x})\mathrm{Tr}_{e}(\mathcal{%
E}_{t+\tau }[\sigma _{0}])+\frac{1-\mathrm{Tr}_{e}(\mathcal{E}_{t+\tau
}[\sigma _{0}])}{d}.  \label{Dosol}
\end{equation}%
On the other hand, consistently with Eqs.~(\ref{RhoEnvPostYX}) and~(\ref%
{P(z|yx)}), $P_{3}(z|x)$ can be written as%
\begin{eqnarray}
P_{3}(z|x) &=&\sum_{y}\Big{\{}\mathrm{Tr}_{s}(E_{z}\rho _{\tau }^{y})\mathrm{%
Tr}_{s}(E_{y}^{t}\rho _{t}^{x})\mathrm{Tr}_{e}(\mathcal{E}_{t+\tau }[\sigma
_{0}])  \notag \\
&&\!\!\!+\sum_{k}\mathrm{Tr}_{s}(E_{z}\rho _{\tau }^{y})\mathrm{Tr}%
_{s}(E_{y}^{t}\mathcal{S}_{k}^{t}[\rho _{t}^{x}])\mathrm{Tr}_{e}(\mathcal{E}%
_{\tau }\mathcal{E}_{t}^{k}[\sigma _{0}])  \notag \\
&&\!\!\!+\sum_{k}\mathrm{Tr}_{s}(E_{z}\mathcal{S}_{k}^{\tau }[\rho _{\tau
}^{y}])\mathrm{Tr}_{s}(E_{y}^{t}\rho _{t}^{x})\mathrm{Tr}_{e}(\mathcal{E}%
_{\tau }^{k}\mathcal{E}_{t}[\sigma _{0}])  \notag \\
&&\!\!\!+\sum_{k}\sum_{j}\mathrm{Tr}_{s}(E_{z}\mathcal{S}_{k}^{\tau }[\rho
_{\tau }^{y}])\mathrm{Tr}_{s}(E_{y}^{t}\mathcal{S}_{j}^{t}[\rho _{t}^{x}]) 
\notag \\
&&\times \mathrm{Tr}_{e}(\mathcal{E}_{\tau }^{k}\mathcal{E}_{t}^{j}[\sigma
_{0}])\Big{\}}.  \label{Tresol}
\end{eqnarray}%
The equality of the first term in Eq.~(\ref{Dosol}) and in Eq.~(\ref{Tresol}%
) is fulfilled due to the DNI condition $\sum_{y}\rho _{\tau }^{y}\mathrm{Tr}%
_{s}(E_{y}^{t}\rho _{t}^{x})=\rho _{t+\tau }^{x}.$ The remaining
contributions can be equated if $\mathrm{Tr}_{e}(\mathcal{E}_{\tau }\mathcal{%
E}_{t}^{k}[\sigma _{0}]),$ $\mathrm{Tr}_{e}(\mathcal{E}_{\tau }^{k}\mathcal{E%
}_{t}[\sigma _{0}]),$ and $\mathrm{Tr}_{e}(\mathcal{E}_{\tau }^{k}\mathcal{E}%
_{t}^{j}[\sigma _{0}])$ do not depend on indexes $k$ and $l,$ which due to
the relation~(\ref{HWeyl}) lead to the condition%
\begin{eqnarray}
1-\mathrm{Tr}_{e}(\mathcal{E}_{t+\tau }[\sigma _{0}]) &=&\mathrm{Tr}_{e}(%
\mathcal{E}_{\tau }\mathcal{E}_{t}^{k}[\sigma _{0}])+\mathrm{Tr}_{e}(%
\mathcal{E}_{\tau }^{k}\mathcal{E}_{t}[\sigma _{0}])  \notag \\
&&+\mathrm{Tr}_{e}(\mathcal{E}_{\tau }^{k}\mathcal{E}_{t}^{j}[\sigma _{0}]),
\end{eqnarray}%
that is, Eq.~(\ref{ConditionalDiscoPresente}).


\begin{thebibliography}{99}
\bibitem{tripartito} This kind of evolutions always admit an underlying
(tripartite) microscopic Hamiltonian description, which includes extra
degrees of freedom that lead to the bipartite dynamics. Nevertheless, in
similitude with the derivation of Markovian quantum master equations~\cite%
{breuerbook}, an effective description through a (non-unitary bipartite)
Lindblad equation applies when the time-scales where unitary effects could
be observed are experimentally or intrinsically unattainable.
\end{thebibliography}
\end{document}